\documentclass[onecolumn]{aa}
\usepackage{graphicx}

\begin{document}
\title{Properties of radio pulses from lunar EeV neutrino showers}
\author{A. R. Beresnyak\inst{1}} 
\offprints{A. R. Beresnyak}
\institute{Puschino Radio Astronomy Observatory, Puschino,\\
Moscow region 142290, Russia \\
e-mail: beres@prao.psn.ru}
\date{\ }

\titlerunning{Properties of radio pulses from neutrino showers}
\authorrunning{A. R. Beresnyak}

\def\om{\omega}


\abstract{
We present results of the simulation of the intensity distribution
of radio pulses from the Moon due to interaction of EeV neutrinos
with lunar regolith. The radiation mechanism is of coherent \^Cerenkov
radiation of the negative charge excess in the shower, known as
Askar'yan effect. Several realistic observational setups with
ground radio telescopes are considered. Effective detector volume
is calculated using maximum-knowledge Monte Carlo code, and the
possibilities to set limits on the diffuse neutrino flux are discussed.
}

\maketitle
\section{Introduction}
Due to the lack of atmosphere on the Moon, its surface
has been proposed as a target for detection of cosmic rays
by mounting detectors on the Moon as early as in the
original Askar'yan's papers (Askaryan, 1962, 1965). Using the whole
visible lunar
surface with its huge effective volume by monitoring the Moon with
ground based radiotelescope has been first proposed by Dagkesamanskii
\& Zheleznykh, 1989. 
Large effective volume presumed detection of particles with
extremely low fluxes, while the distance from the observer limits
to the very energetic events. Using Askar'yan's effect for detection
of neutrinos of cosmic rays is considered advantageous, since shower
coherent radio emission grows roughly as a square of initial particle
energy, and with this consideration in mind it had hoped to overcome the
steep decline in flux expected from known CR spectrum. However
the first rough estimates (Dagkesamanskii
\& Zheleznykh, 1989), has been shown to be too optimistic.
This is due to the particular target geometry: neutrinos at EeV energies
are expected to have rather high cross-section, and
the detectable events will be only those crossing the edge of the Moon,
and the fact that \^Cerenkov angle is complimentary to the full internal
reflection angle.

Up to now, there has been several attempts to detect ultrashort radio
pulses from the Moon (Hankins et al. 1996; Gorham et al, 2001)
none of which shown any signs of such
pulses. However, these observations, combined with the detailed modeling
of the emission will allow to reject particular models of EeV neutrino
spectrum.

Particles with energies around $10^{20}$eV and higher became one
of the major interest in the field of CR science since the formulation
of the GZK paradox (Greisen et al, 1966; Zatsepin \& Kuzmin, 1966).
That is, starting with $5\cdot10^{19}$eV particles
loose energy on the pion production interacting with CMB on scales of around
10Mpc, while the our Galaxy's or stochastic intergalactic magnetic fields
is not high enough to sufficiently curve them. So we expect to see
the sources of such particles which should be relatively nearby, but up
to now, with around a hundred cosmic ray events detected near the GZK energy,
they seem to be distributed rather uniform on the sky.

The expected sources of neutrinos with GZK energies, ranging from
certain, such as GZK cosmic rays, to possible, such as AGNs
and exotic -- topological defects or massive relic particles, predict
neutrino flux somewhat or of the order of magnitude higher than CR flux.
Beyond GZK limit this difference might be even bigger,
so it is no wonder that detection of extra high energy neutrinos received
a considerable attention lately (Alvarez-Mu\~niz et al, 2000;
Razzaque et al, 2002; Provorov \& Zheleznykh, 1995; Buniy \& Ralston, 2002;
Alvarez-Mu\~niz \& Zas, 1997, 1998; Gandhi et al, 1998).

The observation of the Moon with ground radio telescopes could either
detect EeV neutrinos or put a limit on their flux. This paper studies
various aspects of such observations and tries to find the most favorable
setups.

\section{Coherent \^Cerenkov radiation}

Well known Frank-Tamm formula for particle \^Cerenkov losses in dense media,
usually found in textbooks is not applicable in our case. This formula
addresses the case of infinite track length, hence, inevitably,
near-field region. However, the formula for finite track, in the
Fraunhofer limit, has indeed been known for a long time (Tamm, 1939):

$$\frac{d^2P}{d\omega d\Omega}=\frac{ne^2\omega^2L^2}{4\pi^2c^3}
\sin^2\theta\frac{\sin^2X}{X^2}, \eqno (1)$$

Here $P$ is the {\it one-sided} energy spectral density radiated in a given
solid angle, $n$ is the refraction index, $L$ is the track length,
$\theta_c$ is the \^Cerenkov angle, $X$ is the phase equaled to
 $n\omega L(\cos\theta_c-\cos\theta)$
\footnote{CGS units are used unless otherwise noted.}.
The most prominent difference from the infinite track formula is
in the frequency
dependence with flux growing like the frequency squared, rather than the
frequency in the first power.

Maxwell's equation for the infinite space filled with dielectric
with permittivity of $\epsilon$, and permeability of unity, will give
the following solution for the Fourier component of the electric field

$$ \vec E(\omega,\vec x)=\frac{i\omega}{c^2}\int dt'\,d^3\vec x' 
e^{i\om t'+ik|\vec x- \vec x'|} \frac{\vec J_\perp}{|\vec x- \vec x'|}. \eqno (2)$$

Here we defined Fourier component of $\vec E$ as

$$ \vec E(\omega)=\int\limits_{-\infty}^{\infty}\vec E(t) e^{i\om t}\,dt. \eqno (3) $$

In this normalization the radiated energy spectral density, similar to one presented in (1), one-sided, with negative and positive frequency parts summed will be

$$ \frac{d^2P}{d\omega d\Omega}=\frac{cn}{4\pi^2}|R E(\om)|^2. \eqno (4) $$

From the first glance (2) does not contain any traces of the
properties of the media. However, $k$ here is not a variable, it is
determined by the \^Cerenkov resonance condition $k=n\om/c$. 
This expression can be reduced further by using Fraunhofer, or far field
limit, that is, approximating $|\vec x- \vec x'|$ to the linear order and
assuming a particular charge current distribution.

Since particles in a shower move mostly along one particular axis, the
current density could be decently approximated by

$$\vec J_\perp(\vec x, t)=Q(z)\hat n_\perp c\sin\theta\, \delta^3(\vec x- n_zct). \eqno (5) $$

And the electric field, from (2), 

$$ \vec E(\omega,\vec x)=\frac{i\omega}{c^2}\sin\theta\frac{e^{ikR}}{R}
n_\perp\int Q(z')e^{ipz'}\,dz', \eqno (6)  $$
where $p=(1-n\cos\theta)\omega/c$. The angular dependence of the electric
field amplitude near \^Cerenkov angle may be seen as a Fourier transform
of the longitudinal charge distribution. From here it is easy to reproduce
finite track Frank-Tamm formula, (1).

This so called one-dimensional approximation
works surprisingly well in describing maximum intensity at low frequencies
and angular dependence (Alvarez-Mu\~niz et al, 2000). But, rather obviously,
it fails to describe
decoherence that comes from the transverse spread of the shower.

There has been several independent Monte-Carlo simulations of showers
and their \^Cerenkov electric field (Zas et al, 1992; Razzaque et al, 2002;
Provorov \& Zheleznykh, 1995). The results are in
pretty good agreement, concerning integral quantities
such as the total projected (e-p) tracklength, $L$, which determines the field
at the \^Cerenkov angle at low frequencies.

$$ R|E(\omega,\theta=\theta_C)|=\frac{e\om L\sin\theta}{c^2} 
\eqno (7) $$

The tracklength has been shown to scale linearly with the energy
of the showers, or with the electromagnetic energy in the hadronic showers.
The accelerator observation of the effect (Saltzberg et al, 2001) does also
comply reasonably well with this simulations.

The first estimates of the decoherence from the transverse spread of
the shower were done as early as in one of the pioneering Askaryan
papers (Askaryan, 1965). Indeed, this issue is important for choosing the best
observation frequency and for estimates of maximum coherent flux.

In the present paper we adhere to the particular parametrization for the
electric field of one of these simulations, (Alvarez-Mu\~niz et al, 2000),
assuming that
tracklength, decoherence frequency and the angular width of the
\^Cerenkov cone scale with radiation length.

The decoherence is described by the phenomenological form-factor
fitting electric field intensity at the \^Cerenkov angle,
(Alvarez-Mu\~niz et al, 2000),

$$ F(\nu)=\frac{1}{1+(\nu/\nu_0)^{1.44}}, \eqno (8) $$

We have to note, however, that different shower MC's do not agree very
well at frequencies higher than decoherence frequency, which is around
$2.5$GHz for regolith. 

\section{Showers}

As we see from (6) the longitudinal profile of the shower determines
the angular width and the form of the \^Cerenkov electric field peak.
Showers with energies higher than 1~TeV but less than 1~PeV do not fluctuate
much and have profiles close to gaussian, with longitudinal spread growing
slowly
with energy as $\sqrt{2/3\log(E_0/E)}$ (Rossi, 1952). Starting with
energies around 10~PeV, electromagnetic showers are affected
by Landau Pomeranchuk Migdal effect (Landau \& Pomeranchuk, 1953;
Migdal, 1956, 1957). Due to suppressed interaction
at high energies the shower looks as a combination of several subshowers,
with normal length, that is of several radiation lengths. The number
of these subshowers are usually not enough to add up to the good averaged
gaussian profile, so LPM showers fluctuate a lot. In fact, each of them
has the individual form. However the Fourier transform of such a sparse
showers have some general features. Namely, there are two characteristic
lengths, one of the total length of the shower, and the other
is 7-9 radiation lengths, the typical length of a non-LPM shower at these
energies. So, the Fourier transform have an envelope inside of which
there is a typical interference pattern, and the total shower length
determine the width of an individual peaks.

Neutrino or antineutrino interacting with nucleon's quark by charged
current ($W^{\pm}$) gives the corresponding charged lepton and changes
the flavour of the quark. The charged lepton, say, the electron, then
initiate electromagnetic shower of energy $(1-y)E_\nu$, and the quark
initiate hadronic shower of energy $yE_\nu$. Due to the high multiplicity
of the collisions governed by strong force high initial energy in hadronic
shower got divided pretty quickly, thus mitigating the LPM effect.
It has been shown than some hadronic showers of energy as high as 100EeV
show little or no LPM elongation, while the others are lengthening only by
several tenths percent (Alvarez-Mu\~niz \& Zas, 1998). So, the charge
current interacting neutrino produces two distinctly different showers, 
in the case of initial electron neutrino they are develop in the same place
and the \^Cerenkov electric field will be the sum of that of each of them.
The neutral current ($Z$) interactions got neutrino scattered, and
produces only hadronic shower of energy of $yE_\nu$.

The deep inelastic scattering (DIS) cross sections for the energies in
question is determined mainly by the behavior of the distribution function
of the sea quarks at low values of $x$, such as $x\approx M_W^2/2ME_\nu y$
which is $3\cdot10^{-8}/y$ at $E_\nu=10^{20}\mbox{eV}$. The distribution
function at these $x$ has not been measured directly, however it could
be extrapolated from lower values of $x$ by the power law with index
of around -1.3 (Gandhi et al. 1998). We assumed $xq_s(x)\approx x^{-0.33}$.
This
parametrization gives an semianalytic expression for the $d\sigma/dy$
dependence we are interested in. For example, for muon neutrino the
differential cross section is proportional to $(3+2(1-y)^2)y^{-0.67}$
for charged current and $(1+(1-y)^2)y^{-0.67}$ for neutral current,
mean values of y being 0.2 and 0.19 correspondingly.

In our simulation the total $\nu N$ CC and NC cross sections were taken from the parametrization in (Gandhi et al. 1998), $\sigma_{NC}$ being a
0.42 fraction of $\sigma_{CC}$.
 
\section{Time dependence}

Determining the time dependence of the real \^Cherenkov pulse might be a
tricky matter. Even though early papers on Askaryan effect suggested using
full-bandwidth ionospheric-delay compensated radio observations which will
maximize signal-to-noise ratio and make use of the unique \^Cherenkov pulse
shape (Dagkesamanskii \& Zheleznykh, 1989), it would probably be the way only
future dedicated instruments will work. All experiments up to date used
a bandwidth limited, maximum flux triggered setup.

It was suggested to use the frequency domain electric field strength
directly from the theory (Gorham et al, 2001), as the receiving system
works with voltages, which are basically the electric field times the antenna
effective height. However, certain disadvantages come with this approach.
The different simulations of showers often use different fourier transform
normalisation, which sometimes lead to confusion (Razzaque et al, 2002).
In the real experiment voltage is always recorded in the time domain.
Also, the effective height of the antenna is not always easy to get,
and the effective area may change due to orientation. In radio astronomy
it is more common to calibrate receiving system with one or two strong
radio sources. This calibration relates mean squared voltages with the
spectral flux of the source. It is also easier to speak of decoherent
weakening of the signal, such as scattering, in terms of the power, rather
that the amplitude.

In a maximum flux triggered setups, used in today (GLUE) and proposed
for future experiments (ANITA), the phase characteristic of the system
plays a major role, while the radio astronomers are used to fluxes
and usually cut several narrow bands in the primary wide band. For a gaussian
characteristic of the system without phase shifts the maximum flux recorded
will be the one-sided energy spectral density times
$2\pi(\Delta f)_{1/2}/1.18$, in the assumption that the signal phase and
amplitude do not change significantly over this band. 
For filters with flat top used in radio receiving systems this coefficient
is lower, but more impotantly, it is hard to keep zero phase shifts over
a wide bandwidth. In general, we can use the {\it effective response time}
of the system, which is the maximum recorded flux divided by the energy
spectral density of the delta-like pulse. For real systems this time might
be much larger than that determined by uncertainty relation
($1/(2\pi\Delta f)$), thus making maximum flux trigger setup less effective.
The calibration of the system with sufficiently short pulses is almost always
neccesary for this setup.

\section{Geometry and refraction}

In this section, and later, in Monte Carlo, we limit ourselves with
purely refracted pulses. The reason for this is that only refracted
pulses, as opposed to scattered over surface roughness ones,
keep their unique ultra-short time structure. The typical timescale
of a pure pulse which is determined
by the  decoherence frequency 2.5 GHz is around 60 picoseconds, while
the delays from the radiation scattered over the range of several tens
of meters would be hundreds of nanoseconds. Thus, the amplitude and phase
characteristics of the scattered pulses are affected down to frequencies
of 100 MHz or even lower.

The solid angle of the part of the ray caught by the telescope is much
smaller than the typical refracted \^Cerenkov cone, 
so we can use ray optics. According to the effect mentioned in the
introduction there should be strong damping of the outgoing radiation,
however, as it was found, the transmissivity at the regolith-vacuum
interface is not much to blame. As it follows from the geometry of the
most detectable events, the $E$ vector will lay close to the plane of
refraction, however the $T_\|$ transmission coefficient, being zero at
total reflection angle, rises sharply to 1 in a few degrees.

It is the solid angle stretching near zero angles that does the most
effect. Indeed, the solid angle before and after refraction will
have the ratio of

$$\frac{d\Omega}{d\Omega'}=\frac{\sin\beta}{n^2\sin\alpha}, $$

where $\beta$ is an angle between the plane of the surface of the Moon
and the outgoing ray, and $\alpha$ is the same angle for incoming ray.
This coefficient scales like $\beta$ at small angles, as well as $T_\|$ does,
however it is much smaller than $T_\|$ for the angles of a few degrees
-- the ones we are interested in. As an example in Figure 1 we showed
the full exact damping coefficient which is the product of the
above ratio to the $T_\|$, plotted versus $r=1-\cos\beta$. 
$r$ may be thought of as a relative distance of the event from lunar limb,
$r=0$ correspond to the limb, $r=1$ to the center.
Both transmissivity and solid angle stretching is proportional to $\beta$ at
small $\beta$, however for transmissivity this law fails already at a few
degrees.

\begin{figure}
\caption{The full damping at the regolith-vacuum interface, versus
relative distance from the limb}
\includegraphics{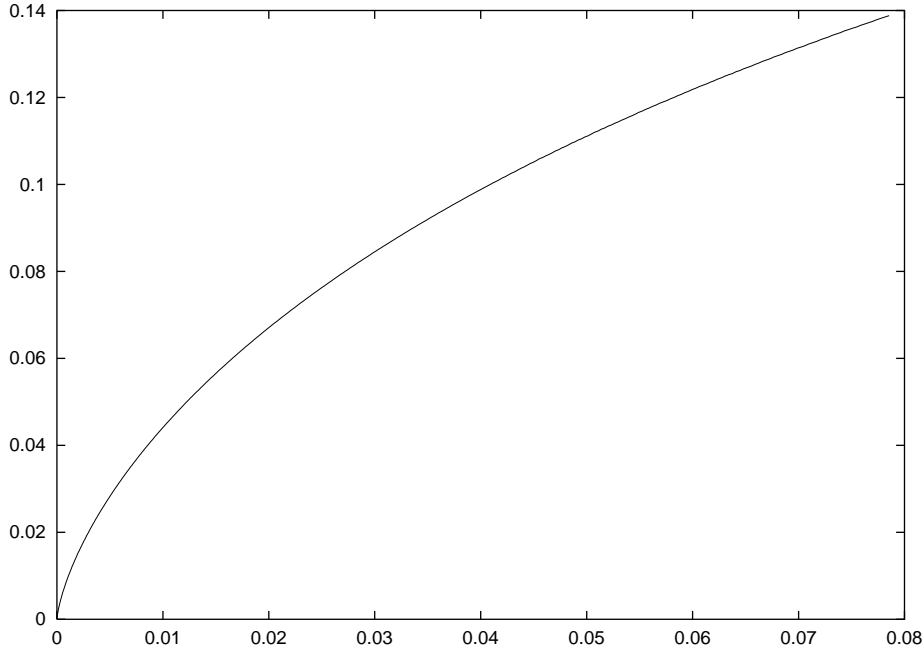} 
\end{figure}

It is also interesting to calculate which part of the incoming particles
full $4\pi$ solid angle will give radiation coming out of the Moon, though
not necessarily detectable, in the assumption that all radiation is
concentrated in the cone with typical angle estimated in section 2.
This ``effective'' solid angle is shown on figure 2.
It happened, that those neutrinos that have positive entrance angles,
that is those, that hit the visible hemisphere, give important
income to this total angle, especially at low frequencies and low $\beta$'s.

As we see from Figures 1 and 2 both the effective solid angle and the
intensity damped by more than order of magnitude, which undermines
previous optimistic estimates (Dagkesamanskii \& Zheleznykh, 1989) and
brings close attention to the
detailed modeling of the transmissivity properties of the lunar-vacuum
interface. In the next section we describe such a modeling.
 
\begin{figure}
\caption{The effective solid angle versus output angle, in the
assumption that radiation coming from \^Cerenkov cone is always observed}
\includegraphics{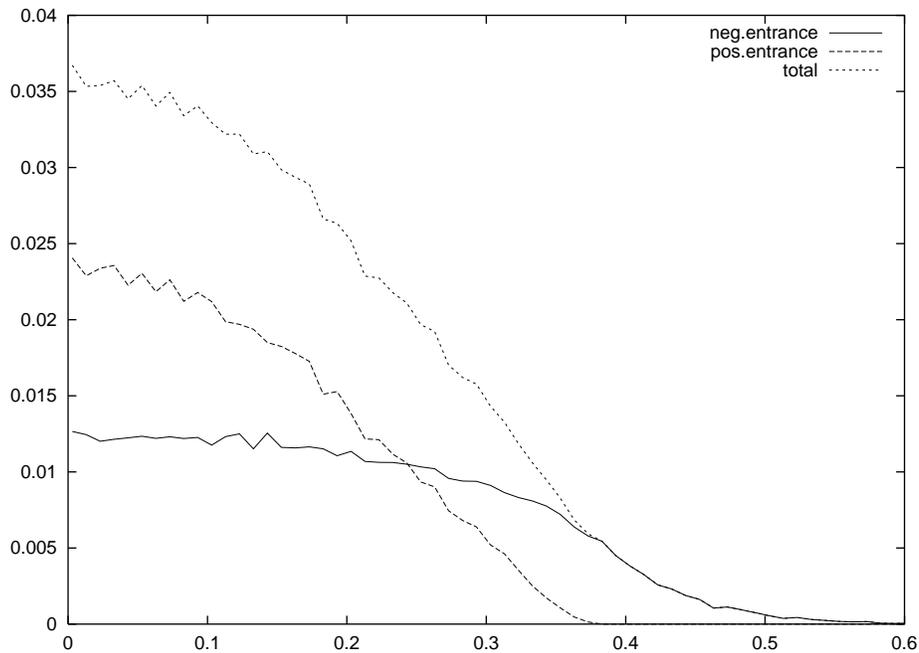} 
\end{figure}

\section{Monte Carlo}

We modeled a 64 meter telescope with standard circular aperture, pointed
at some point between lunar center and the limb, making observations
at some frequency $\nu$ with a bandwidth of 100 MHz, set up so to
trigger events which have a total flux density higher than a certain
threshold.

For each prospective CC or NC event we generated its lunar coordinates,
depth at which interaction took place, two angles of the incoming neutrino,
Bjorken y, calculated the radio flux taking into account LPM by
parametrizations from (Alvarez-Mu\~niz \& Zas, 1997, 1998) and polarization
position angle. We used
exact formulae for transmissivity of two different polarizations,
adopted absorption length of 15~m(1~GHz/$\nu$). We assumed that regolith
or some other material with similar properties cover the Moon with a layer
of at least 30 meters deep, which somewhat increased effective
volume at very high energies. We adopted the lunar surface slopes model
with the gaussian distribution of the slope angles with an rms of of $6^o$
at the length scales we are interested in.
The angular dependence of the shower radio flux, and the angular
dependence of the sensitivity of the telescope were both taken into account.

\begin{figure}
\caption{Some events as seen on the Moon's face. Upper circle
is the Moon's limb, lower shows the telescope beam. Polarization
is shown with electric field vectors.}
\includegraphics{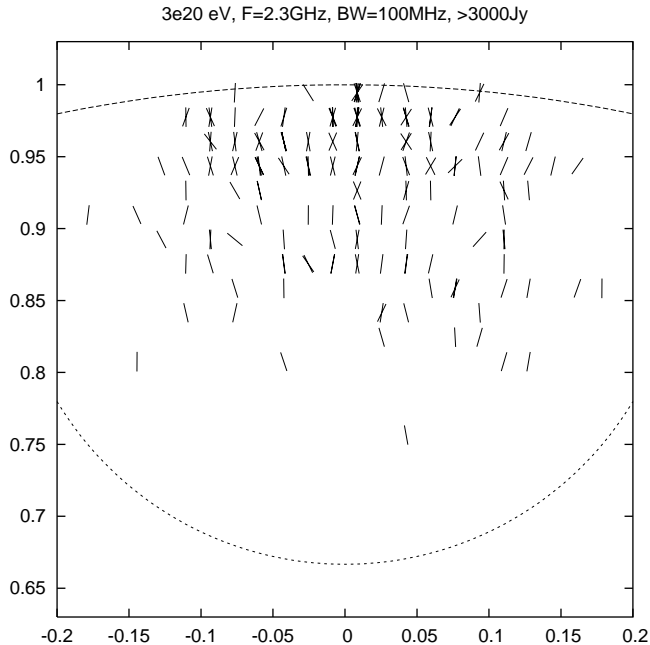} 
\end{figure}

Several events with their polarizations were drawn on Figure 3.
The frequency of observation was 2.3~GHz and the flux density limit 3000~Jy.
The telescope was aimed $13.5'$ from the lunar center. As we see, most of
the events are concentrated close to the rim, and this behavior becomes more
prominent with higher energies, since the neutrino interaction length
shortens. As it comes from a theory, all events are fully linearly
polarized, and polarization tends to align along radius-vector to the
lunar center.

Neutrinos with positive entrance angles give almost no contribution
at low energies, $<10^{20}$eV, however their relative contribution grows
as cross-section grows, at $10^{23}$eV contributing about the half
of all trigger events.

Even though the mean value of $y$ is around 0.2, that is most of the
time most of the energy goes to either electromagnetic shower or neutrino,
in this geometry electromagnetic showers happen to be relatively
unimportant for radio detection. They contribute to trigger significantly
only close to the edge of detection, which is around $10^{20}$eV for
2.3~GHz and a bandwidth of 100~MHz. The number of triggers came purely
from EM showers do not grow much with energy. 

The effective detector aperture for the frequency of 2.3GHz and a limit of
3000~Jy as dependent on primary particle energy is shown on Figure 4. Also
shown apertures from two GLUE publications.

\begin{figure}
\caption{\ }
\includegraphics{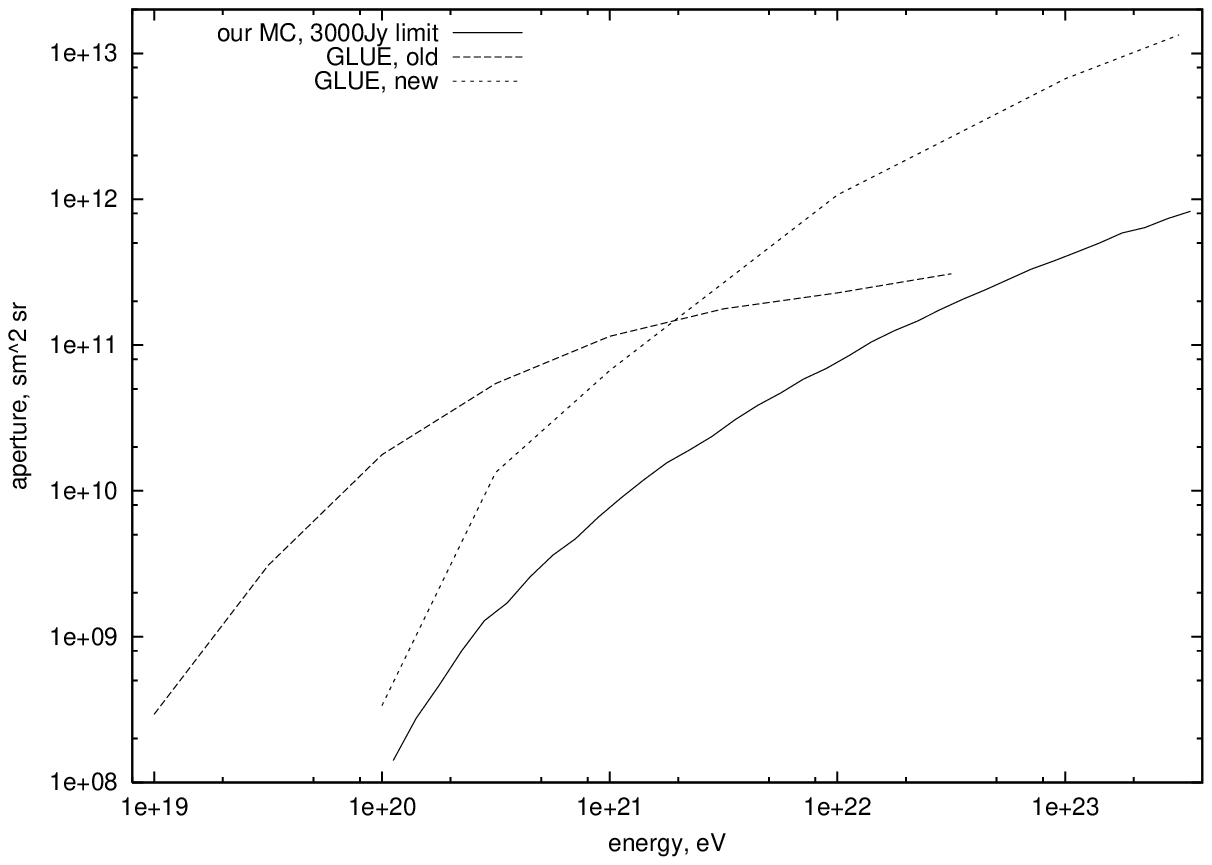} 
\end{figure}

We show effective apertures instead of effective volumes, traditional
in neutrino experiments, for a good reason. In order to provide predictions on
flux we need aperture, which is usually deduced from volume and the cross section. In our case, however, the cross section itself is under discussion, and
the simulation naturally use cross sections to calculate shadowing factors.
Thus, showing apertures we avoid the situation when someone use our effective
volume with the cross section different from that we used.

Even though radio location of the Moon suggest the rms of the slopes to
be around $6^o$, the different parts of the surface has this value different.
Such as the places abundant in craters have significantly higher rms slope
and marines have lower ones. Hence, in the case of a large single dish and
high observation frequency, when the Moon is resolved, it might be
advantageous to look at the places with craters. The present ground
radio techniques, however, can only measure rms of the slopes of the frontal
part of the surface of the Moon, and since we almost always want to aim
at the rim, these results must be taken with caution, until the lunar surface
roughness is explored in more detail. On Figure 5 we show the dependence
of the effective volume, as opposed to the rms of the slope in degrees, for
three neutrino energies.

\begin{figure}
\caption{Effective volume vs rms of the slopes}
\includegraphics{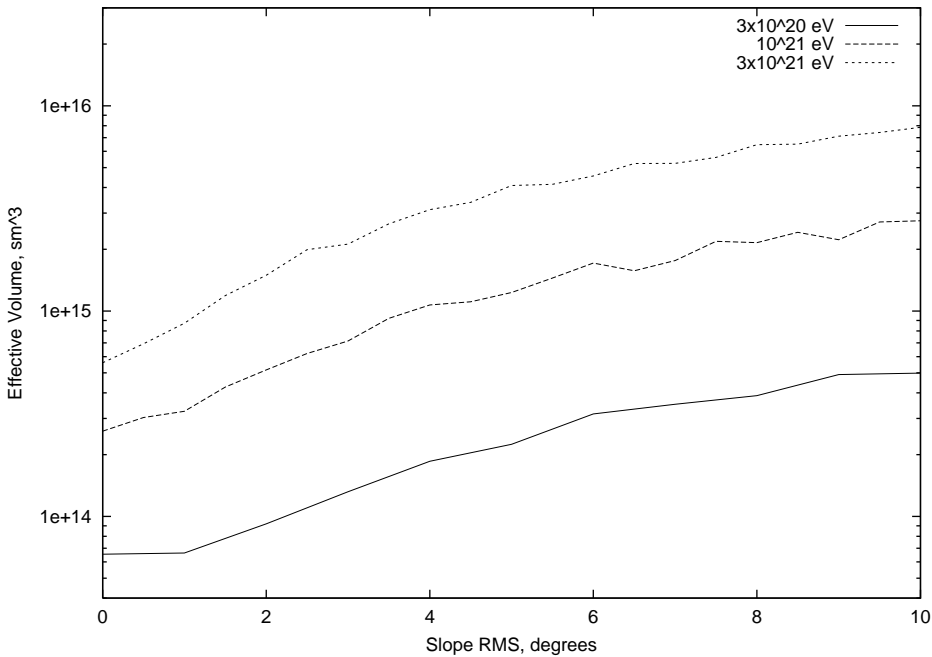} 
\end{figure}

\section{Diffuse neutrino flux limits and discussion}

Having the observational time, and the effective
aperture deduced in a previous section it is straightforward to estimate
a diffuse neutrino flux at a given energy, having several events detected,
or put an upper limit on flux if there were no detection. However, the fact
of no-detection, logically, allows us to reject only one {\it particular}
model spectrum of the cosmic neutrinos. It allows not the upper limits on
a real differential flux curve $EdF/dE$ at any given energy. Indeed,
the differential flux might have a thin but high fluke constituting
only a small amount of an integrated flux, which surely will
not be detected.

For the purpose of the GLUE experiment (Gorham et al, 2001) the so called
model-independent limit of flux is described as the limit
on the the $EdF/dE$ curve equaled inverse of the product of
aperture and the observational time, corrected by the appropriate
Poisson factor for the certain confidence level. This limit is appoximately
true, if the flux does not change significantly over the order of magnitude
of energy.
However, we propose a bit more strict limit, which is multiplied by
the factor of $1/\log(\Delta E)$, thus allowing to rule out the total flux
between and under two ajacent points. The closer points are, the higher
will be the limit.

As we see from previous discussion, especially from figure 4, the lunar
neutrino experiments, such as GLUE, or Pushchino experiment are rigged
with uncertanty in the interpretation. Even the estimates of the aperture
given by the same group changes significantly. In the case of purely refracted
pulses considered in this article, the biggest uncertanty is in the fact, that
we do not know exactly neither the distribution of the slopes, nor the number
of the pulses that will be refracted purely, as opposed by scattered by
small irregularities of the surface.

\begin{acknowledgements}
I am grateful to R.~D.~Dagkesamanskii and I.~M.~Zheleznykh for fruitful
discussion and suggestions. This work was partially supported by
Russian Foundation of Base Research grant N 02-02-17229.
\end{acknowledgements}

\begin{center}
{\Large References}
\end{center}
\parindent0pt

Alvarez-Mu\~niz~J.; V\'azquez~R. A.; Zas~E., 2000, Phys. Rev. {\bf D},{\bf 62},063001

Alvarez-Mu\~niz~J., Zas~E., 1997, Phys.Lett. B411, 218-224

Alvarez-Mu\~niz~J., Zas~E., 1998, Phys.Lett. B434, 396-406

Askaryan~G.A., 1962, JETP 14, 441

Askaryan~G.A., 1965, JETP 21, 658

Buniy~R.V., Ralston~J.P., 2002, Phys.Rev. {\bf D 65}, 016003

Butkevich~A.V., et al, 1988, Z. Phys. C, {\bf 39}, 241-250

Dagkesamanskii~R.D., Zheleznykh~I.M., 1989, JETP Lett, 50, 233

Gandhi~R., et al, 1998, Phys.Rev. D {\bf 58}, 093009

Gorham~P.W., et al, 2001, Proc. RADHEP-2000, p.177.

Greisen~K., 1966, Phys. Rev. Lett. {\bf 16}, 748

Hankins~T.H., Ekers~R.D., O'Sullivan~J.D., 1996, MNRAS, 283, 1027.

Landau~L., Pomeranchuk~I., 1953, {\sl Dokl.\ Akad.\ Nauk\
SSSR} {\bf 92}, 535; {\bf 92}, 735

Migdal~A.~B., 1956, Phys. Rev. {\bf 103}, 1811

Migdal~A.~B., 1957, Sov. Phys. JETP {\bf 5}, 527
 
Provorov~A. L. and Zheleznykh~I. M., 1995, Astropart. Phys. 4, 55

Razzaque~S., et al, 2002, Phys. Rev. {\bf D}, {\bf 65}, 103002

Rossi~B., 1992, High Energy Particles (Prentice Hall, New York)

Saltzberg~D., Gorham~P.W., et al., 2001, Phys. Rev. Lett., {\bf 86}, 13, 2802.

Tamm~I.E., 1939, J. Phys. (Moscow) 1, 439

Zas~E., Halzen~F., and Stanev~T., 1992, Phys. Rev. {\bf D}, {\bf 45}, 362

Zatsepin~G.T., Kuzmin~V.A., 1966, Pisma Zh. Eksp. Teor. Fiz. {\bf 4}, 114

\end{document}